\documentclass{svmult}
\usepackage{graphicx}
\begin{document}
\title*{Observational Cosmology}
\author{R.H. Sanders}
\institute{Kapteyn Astronomical Institute, Groningen, The Netherlands}
\maketitle
\def\ho{$H_o\,\,$}
\def\apj{{\it Astrophys.J.} }
\begin{abstract}
I discuss the classical cosmological tests-- angular size-redshift,
flux-redshift, and galaxy number counts-- in the light of the
cosmology prescribed by the interpretation of the CMB
anisotropies.  The discussion is somewhat of a primer for physicists,
with emphasis upon the possible systematic uncertainties in the
observations and their interpretation.  Given the curious composition
of the Universe inherent in the emerging cosmological model,  I stress 
the value of searching for inconsistencies rather than concordance, and 
suggest that the prevailing mood of triumphalism in cosmology is
premature.
\end{abstract}
\section{Introduction}

     The traditional cosmological tests appear to have been 
overshadowed by observations of the anisotropies in the cosmic microwave
background (CMB).  We are told that these observations accurately
measure the geometry of the Universe, its composition, its present
expansion rate, and the nature and form of the primordial fluctuations
\cite{speal03}.
The resulting values for these basic parameters are very similar to 
those deduced earlier
from a variety of observations-- the so-called ``concordance model''-- 
with about 30\% of the closure density of the Universe comprised of
matter (mostly a pressureless, non-baryonic dark matter), the remainder
being in negative pressure dark energy \cite{osst95}.
Given the certainty and precision of these assertions, 
any current discussion of 
observational cosmology must begin with the question:  Is there any
room for doubt?  Why should we bother with lower precision cosmological
tests when we know all of the answers anyway?

While the interpretation of the CMB
anisotropies has emerged as the single most important cosmological tool,
we must bear in mind that the conclusions drawn do
rest upon a number of assumptions, and the results
are not altogether as robust as we are, at times, led to believe.  
One such assumption, for example, is that of adiabatic initial 
fluctuations-- that is,
100\% adiabatic.  A small admixture of correlated isocurvature
fluctuations, an aspect of braneworld scenarios \cite{maart}, 
can affect 
peak amplitudes and thus, the derived cosmological
parameters.  A more fundamental assumption is that of the validity of 
traditional Friedmann-Robertson-Walker
(FRW) cosmology in the post-decoupling universe.  Is the expansion of the
universe described by the Friedmann equation?  Even minimal changes to
the right-hand-side, such as the equation of state of the dark energy
component, can alter the angular size distance to the last scattering
surface at z=1000 and the luminosity distance to distant supernovae.  
But even more drastic changes to the Friedmann equation,
resulting from modified gravitational physics, have been proposed 
in attempts to remove the unattractive dark energy
\cite{def01,ceal}.

Such suggestions reflect a general unease with the concordance model-- 
a model that
presents us with a universe that is strange in its composition.  The most
abundant form of matter consists of, as yet, undetected non-baryonic particles
originally postulated to solve the problems of structure formation and
of the missing mass in bound gravitational systems such as galaxies and
clusters of galaxies. In this second respect, it is fair to say that it has 
failed-- or, to be generous, not yet succeeded-- because the predicted 
density distribution of dark halos which emerge from cosmic N-body 
\cite{nfw} simulations appears to be 
inconsistent with observations of spiral galaxies \cite{mdb98} or with 
strong lensing in clusters of galaxies \cite{treal}. 

Even more mysterious is the ``dark energy'',
the pervasive homogeneous fluid with a negative pressure which may
be identified with the cosmological constant, the zero-point energy
density of the vacuum.  The problem of this unnaturally low energy
density, $10^{-122}$ in Planck units, is well-known,
as is the cosmic coincidence problem:  why
are we observing the Universe at a time when the cosmological constant
has, fairly recently, become dynamically important \cite{car01}?  
To put it another way, why are the energy densities of matter and
dark energy so comparable at the present epoch?  
This is strange because the density of matter dilutes with
the expanding volume of the Universe while the vacuum energy 
density does not.  It is
this problem which has led to the proposal of dynamic dark energy,
quintessence-- a
dark energy, possibly associated with a light scalar field-- with an
energy density that evolves with cosmic time possibly tracking the
matter energy density \cite {rapeb}.  Here the difficulty is that
the field would generally be expected to have additional observational
consequences-- such as violations of the equivalence principle at some
level, possibly detectable in fifth force experiments \cite{car01}.

For these reasons, it is even more important to pursue cosmological
tests that are independent of the CMB, because one might expect
new physics to appear as observations inconsistent with the concordance model.
In this sense, discord is more interesting than concord; 
to take a Hegelian point of view-- ideas progress through dialectic, not 
through concordance.  It is with this in mind that I will review
observational cosmology with emphasis upon CMB-independent tests.

Below I argue that the 
evolution of the early, pre-recombination universe is  
well-understood and tightly constrained by considerations of primordial
nucleosynthesis.  If one wishes to modify general 
relativity to give deviations from Friedmann expansion, then
such modifications are strongly constrained at early times, at energies
on the order of 1 MeV.  However, cosmological
evolution is much less constrained in the post-recombination universe
where there is room for deviation from standard Friedmann
cosmology and where the more classical tests are relevant.  I will discuss
three of these classical tests: the angular size distance test where
I am obliged to refer to its powerful modern application with respect
to the CMB anisotropies; the 
luminosity distance test and its application to observations 
of distant supernovae; 
and the incremental volume test as revealed by faint
galaxy number counts.  

These classical tests yield results that
are consistent, to lower precision, with the parameters deduced
from the CMB.  While one can make minimal changes to 
standard cosmology, to the equation of state of the dark energy for
example, which yield different cosmological parameters, there is 
no compelling observational reason to do so.  It remains the peculiar 
composition and the extraordinary coincidences embodied by the 
concordance model that call for deeper insight.  
Such motivations for questioning a paradigm
are not unprecedented;  similar worries led to
the inflationary scenario which, unquestionably, has had the dominant impact
on cosmological thought in the past 25 years and which has found
phenomenological support in the recent CMB observations. 

I am not going to discuss cosmological tests based upon
specific models for structure formation, such as the form of the
luminous matter power spectrum \cite{peak} or the
amplitude of the present mass fluctuations \cite{hoek}.
I do not mean to imply that such such tests are 
unimportant, it is only that I restrict myself here to more global
and model-independent tests.  If one 
is considering a possibility as drastic as a modification of 
Friedmann expansion due, possibly, to 
new gravitational physics, then it is tests of the global curvature and
expansion history of the Universe that are primary.

I am also going to refrain, in so far as possible, 
from discussion of theory-- of new 
gravitational physics or of any other sort.  The theoretical
issues presented by dark matter that can only be detected
gravitationally or by an absurdly small but non-zero cosmological 
constant are essentially not problems for the interpretive
astronomer. The primary task is to realistically access the
reliability of conclusions drawn from the observations, and that is 
what I intend to do.

\section{ Astronomy made simple (for physicists)}

I think that it is fair to assume that most of you are physicists, so
I begin by defining some of the units and terminology used
by astronomers.  I do this because much of this terminology is 
arcane for those not in the field.  

First of all there is the peculiar logarithmic scale of flux-- magnitudes--
whereby a factor of 100 in flux is divided into five equal logarithmic 
intervals.  The system is ancient and has its origin in the logarithmic
response of the human eye.  The ratio of the flux of two objects is 
then given by a difference in magnitudes; i.e., 

$$ m_2-m_1 = -2.5\,log(F_2/F_1) \eqno(2.1)$$
where, one will notice, smaller magnitude means larger flux.
The zero-point of this logarithmic scale is set by some standard star
such as Vega.  Because this is related to the flux, and not the
luminosity of an object, it is called the ``apparent'' magnitude.
Distant galaxies have apparent magnitudes, in visible light, of greater than
20, and the galaxies in the Hubble Deep Field, go down to magnitudes of 30.
The magnitude is typically measured over a specified 
wavelength range or color band, such as blue (B), visual (V),
or infrared (K), and these are designated $m_B$, $m_V$, and $m_K$, or 
sometimes just B,V, and K.  This is made more confusing by the fact that
there are several competing photometric systems (or sets of filters)
and conversion between them is not always simple. 

With a particular photometric system one can measure the color of an
astronomical object, expressed as difference in magnitudes in two bands,
or color index; e.g.,
$$B-V = 2.5\,log(F_V/F_B) \eqno(2.2)$$
Here a larger B-V color index means that an object is relatively redder;
a smaller B-V that the object is bluer.  Unlike the
apparent magnitude, this is an intrinsic property of the object.  Or rather,
it is intrinsic once the astronomer corrects the magnitudes in the
various bands to the zero-redshift ($z=0$) frame.  This is called the
``K-correction'' and requires a knowledge of the intrinsic spectral
energy distribution (SED) of the source, be it a galaxy or a distant
supernova.

The luminosity of an object is also an intrinsic property and is usually
expressed by astronomers as an ``absolute'' magnitude.  This is the
apparent magnitude an object would have if it were placed at a standard
distance, taken to be 10 parsecs, i.e. $3\times 10^{17}$ m 
(more on parsecs below).  Because this
distance is small by extragalactic standards the absolute magnitudes of
galaxies turn out to be rather large negative numbers: $M_G \approx$
-18 to -21.
The luminosity of a galaxy $L_G$ in units of the solar luminosity
$L_\odot$ can be determined from the relation
$$M_G -M_\odot = -2.5\,log(L_G/L_\odot) \eqno(2.3)$$
where the absolute magnitude of the sun (in the V band) is 5.5.
The luminosities of galaxies typically range from $10^8$ to $10^{11}$
$L_\odot$.  The peak absolute magnitude of a type I
supernova (SNIa) is about -19.5, or comparable to an entire galaxy.  This
is one reason why these objects are such ideal extragalactic distance
probes.

The unit of distance used by astronomers is also archaic:  the parsec which
is about $3\times 10^{16}$ m or about 3 light years.  This is the distance
to a star with an semi-annual parallax of 1 arc second and is not a bad
unit when one is discussing the very local region of the galaxy.
Our galaxy has a diameter between 10 to 20 kiloparsecs, so the kiloparsec
is an appropriate unit when discussing galactic structure.  The appropriate
unit of extragalactic distance, however, is the ``megaparsec'' or Mpc, with
nearby galaxies being those at distances less than 10 Mpc.  The nearest
large cluster of galaxies, the Virgo cluster, is at a distance of 20 Mpc,
and very distant galaxies are those further than 100 Mpc, although here one
has to be careful about how distance is operationally defined.

We all know that the Universe is uniformly expanding and the Hubble 
parameter, $H$, is the recession velocity of galaxies per unit distance,
with $H_o$ being its value in the present Universe.  It is typically
measured in units of km ${\rm s^{-1}{Mpc}^{-1}}$ or inverse time.
A number of observations point to
$H_o \approx 70$ km ${\rm s^{-1}{Mpc}^{-1}}$ .
The Hubble time is defined as $t_H = {H_o}^{-1}$ which is about
$9.8 \times 10^{9}$ $h^{-1}$ years, 
and this must be comparable to the age of the Universe.  The definition
$h=H_o/100$ km ${\rm s^{-1}{Mpc}^{-1}}$  
is a relic of the recent past when the 
Hubble parameter was less precisely determined, 
but I keep using it below because it remains convenient as a unit-less
quantity.
We can also define a characteristic scale for the universe which is
the Hubble radius or $r_H = c/H_o$ and this is 3000 $h^{-1}$ Mpc.  
This would be comparable to the ``distance'' to the horizon.

Just for interest, one could also define a Hubble acceleration or
$a_H=cH_o\approx 7\times 10^{-10}$ m/s$^2$.  This modest acceleration
of 7 angstroms/second squared is, in effect, the acceleration
of the Hubble flow at the horizon if we live in a Universe dominated by a 
cosmological constant as observations seem to suggest. It is also 
comparable to the
acceleration in the outer parts of galaxies where the need for dark
matter first becomes apparent \cite{milg}. In some sense,
it is remarkable that such a small acceleration has led to a major
paradigm shift.

\section{ Basics of FRW cosmology}

The fundamental assumption underlying the construction of cosmological
models is that of the cosmological principle:  The Universe appears 
spatially isotropic in all its properties to all observers.  The only
metric which is consistent with this principle is the Robertson-Walker
metric:
$$ds^2 = c^2dt^2 - {{a^2(t)dr^2\over{[1-r^2/{R_o}^2]}} -
 a^2(t)r^2({d\theta^2} + sin^2(\theta){d\phi}^2)} \eqno(3.1)$$
where $r$ is the radial comoving coordinate, $a(t)$ is the dimensionless
scale factor by which all distances vary as a function of cosmic time,
and ${R_o}^{-2}$ is a parameter with dimensions of inverse length squared
that describes the curvature of the Universe and may be positive, zero, 
or negative (see \cite{wrind} for a general discussion). 

This is the geometry of the Universe, but dynamics is provided by
General Relativity-- the Einstein field equations-- which yield
ordinary differential equations for $a(t)$.  The time-time component
leads to a second order equation:
$$ \ddot{a} = -{{4\pi G}\over 3}a(\rho+3p/c^2) \eqno(3.2)$$
where $\rho$ is the density, $p$ is the pressure and the quantity 
in parenthesis is the active gravitational mass density.
Considering conservation of energy for a perfect fluid 
$$d(\rho V) = {-p{dV}/{c^2}} \eqno(3.3)$$
with an equation of state
$$p = w\rho c^2\eqno(3.4)$$
we have $\rho\propto a^{-1(1+w)}$.  The equation of state combined with
eq.\ 3.2 tells us that the Universe is accelerating if $w<-1/3$.

The space-space components combined with the time-time component yield 
the usual first-order Friedmann equation
$$\Bigl({H\over{H_o}}\Bigr)^2 - {{\Omega_k}\over{a^2}} = 
\sum_i{\Omega_i a^{-3(1+w_i)}} \eqno(3.5)$$
where $H=\dot{a}/a$ is the running Hubble parameter, 
the summation is over the various fluids comprising the Universe and
$$\Omega_i = {{8\pi G\rho_i\over{3{H_o}^2}}}\eqno(3.6)$$ with 
$\Omega_k=-(r_H/R_o)^2$.
We often see eq.\ 3.5 written in terms of redshift where $a = (1+z)^{-1}$.
Each component has its own equation of state parameter, $w_i$: 
$w=0$ for non-relativistic matter (baryons, CDM); $w=1/3$ for radiation
or other relativistic fluid;
$w=-1$ for a cosmological constant; and  
$-1<w<-1/3$ for ``quintessence'', dynamic dark energy resulting in 
ultimate acceleration of the universal expansion.
I will not consider $w<-1$  which has been termed
``phantom'' dark energy \cite{cald}; 
here the effective density increases as the Universe
expands (this could be realized by a ghost field, a scalar with a kinetic 
term in the Lagrangian having the wrong sign so it rolls up rather than down 
a potential hill).

Given a universe composed of radiation, non-relativistic matter,
and quintessence, the Friedmann equation takes its familiar form:
$$\Bigl({H\over{H_o}}\Bigr)^2 - {\Omega_k\over{a^2}} = \Omega_ra^{-4} + 
\Omega_ma^{-3} +\Omega_Q a^{-3(1+w)}. \eqno(3.7)$$ 
Here it is evident that radiation drives the expansion at early times
($a<<1$), non-relativistic matter at later times, a non-vanishing curvature
($\Omega_k\neq 0$) at later times still, and, if $w<-1/3$, 
the vacuum energy density ultimately dominates.  For the purpose of this
lecture, I refer to eq.\ 3.7 with $w=-1$ (the usual cosmological constant)
as standard FRW cosmology,
while $0>w\neq -1$ would represent a minimal modification to FRW
cosmology.  Moreover, when $w=-1$, I replace $\Omega_Q$ by $\Omega_\Lambda$.
I will not consider changes to the
Friedmann equation which might result from modified gravitational physics.

Because the subject here is observational cosmology we must discuss the
operational definitions of distance in an FRW Universe.  If there exists
a standard meter stick, an object with a known 
fixed linear size
$d$ which does not evolve with cosmic time, then one could obviously
define an angular size distance:
$$D_A = {d\over\theta}\eqno(3.8$$
where $\theta$ would be the observed angle subtended by this object.
If there exists a standard candle, an object with a 
known fixed luminosity $L$ which does not vary with cosmic time, then
one could also define a luminosity distance:
$$D_L = \bigl({L\over{4\pi F}}\bigr)^{1\over2}\eqno(3.9)$$
where F is the measured flux of radiation.

For a RW universe both the angular size distance and the luminosity distance 
are related to the radial comoving coordinate,
$$r = |R_o|\chi\Bigl[{r_H\over{|R_o|}}\int_{\tau_o}^{\tau}
{d\tau\over{a(\tau)}}\Bigr] \eqno(3.10)$$
where $\tau=tH_o$, ${R_o}^2=-{r_H}^2/\Omega_k$, and
$$\chi(x) = sin(x)\,\,\,\,\,{\rm if}\,\,\Omega_k<0$$
$$\chi(x) = sinh(x)\,\,\,\,\,{\rm if}\,\,\Omega_k>0$$
$$\chi(x) = x\,\,\,\,\,\,\,\,\,{\rm if}\,\,\Omega_k=0.$$
Then it is the case that 
$$D_A = r\,a(\tau) = r/(1+z) \eqno(3.11a)$$
and
$$D_L = r/a(\tau) = r(1+z). \eqno(3.11b)$$
It is evident that both the angular size distance and the luminosity distance
depend upon the expansion history (through $\int{{d\tau}/a(\tau)}$) and
the curvature (through $\chi(x)$).

The same is true of a comoving volume element:
$$dV = {{r^2}} drd\Omega\eqno(3.12)$$
where here $d\Omega$ is an incremental solid angle.  Therefore, if there
exists a class of objects with a non-evolving comoving density, then
this leads to another possible cosmological test:  simply count those
objects as a function of redshift or flux.

Below, I am going to consider these measures of distance and volume in
the form of three classical cosmological tests:

1.  Angular size tests which essentially involve the determination of
$D_A(z)$.  Here one measure $\theta$ for objects with a known and
(hopefully) standard linear size (such as compact radio sources).

2.  Luminosity distance tests which involve the measurement of $F(z)$ for
presumably standard candles (such as supernova type Ia, SNIa).

3.  dV/dz test which involve the counts of very faint galaxies as a 
function of flux and redshift.

But before I come to these classic tests, I want to discuss the
evidence supporting the validity of the standard hot Big Bang,
as an appropriate description of the early pre-recombination Universe.

\section{ Observational support for the standard model of the 
early Universe}

The discovery 40 years ago of the cosmic microwave background radiation (CMB) 
ended, for most people, the old debate about Steady-State vs.\ the Hot Big
Bang.  Ten years ago, support for the Hot Big Bang was fortified by the
COBE satellite which demonstrated that the CMB has a Planck spectrum
to extremely high precision;  it is, quite literally, the most perfect
black body observed in nature \cite{math}.  This makes any model in which the
CMB is produced by some secondary process, such as thermal re-radiation of
starlight by hot dust, seem extremely difficult, if not impossible, to
contrive.

Not only does the background radiation have a thermal spectrum, it is now
evident that this radiation was
hotter in the past than now as expected for adiabatic expansion
of the Universe.  This is verified by observations of neutral
carbon fine structure lines as well as molecular hydrogen rotational
transitions in absorption line systems in the spectra of distant 
quasars.  Here, the implied population of different levels, determined
primarily by the background radiation field, is an effective thermometer
for that radiation field.
One example is provided by a quasar with an absorption line system at 
z = 3.025 which
demonstrates that the temperature of the CMB at this redshift was
$12.1^{+1.7}_{-8.2}$ K, consistent with expectations ($T\propto 1+z$)
\cite{mol}.

However, the most outstanding success story for the Hot Big Bang is generally
considered to be that of Big Bang Nucleosynthesis (BBN) 
which, for a given number of relativistic particle species, 
predicts the primordial abundances of the light isotopes with, 
effectively, one free parameter:  the ratio of baryons-to-photons, $\eta$
\cite{steig}.
I want to review this success story, and point out that there remains
one evident inconsistency which may be entirely observational, but which
alternatively may point to new physics.

We saw above in the Friedmann equation (eq. 3.7) that radiation, if present,
will always dominate the expansion of the Universe at early enough epochs
(roughly at $z\approx 2\times 10^4 {\Omega_m}$.)
This makes the expansion and thermal history of the Universe particularly
simple during this period.
The Friedmann equation becomes
$$H^2 = {{4\pi G aT^4N(T)}\over{3c^2}}; \eqno(4.1)$$
here $a$ is the radiation constant and $N(T)$ is the number of
degrees of freedom in relativistic particles.   
The scale factor is seen to
grow as $t^{1/2}$ which means that the age of the Universe is given by
$t=1/2H$.  This implies, from eq.\ 4.1, an age-temperature relation of
the form $t \propto T^{-2}$.  Putting in numbers, the
precise relation is
$$t = {2.5\over {{T_{MeV}}^2}{N(T)}^{1\over2}}\, {\rm s} \eqno(4.2)$$
where the age is given in seconds and $T_{MeV}$ is the temperature 
measured in MeV.  It is only necessary to count the number of
relativistic particle species: 
$$N(T) = \sum{g_B} + {7\over 8}\sum{g_F} \eqno(4.3)$$
where the sums are over the number of bosonic degrees of freedom ($g_B$)
and fermionic degrees of freedom ($g_F$).  The factor 7/8 is due to
the difference in Bose-Einstein and Fermi-Dirac statistics.
Adding in all the known species-- photons, electrons-positrons 
(when $T_{MeV}>0.5$), three types of neutrinos and
anti-neutrinos-- we find
$$t \approx {T_{MeV}}^{-2}\, {\rm s}\eqno(4.4)$$
for the age-temperature relation in the early Universe.

When the Universe is less than one second old ($T > 1$ MeV) the weak
interactions 
$$p + e^- \leftrightarrow n+\nu_e$$
$$n+e^+ \leftrightarrow p+ \nu_e$$
$$n\leftrightarrow p+e^-+ \nu_e$$
are rapid enough to establish equilibrium between these various species.
But when T falls below 1 MeV, the reaction rates become slower than the
expansion rate of the Universe, and neutrons ``freeze out''-- they fall
out of thermal equilibrium, as do the neutrinos.
This means the equilibrium ratio of neutrons to protons at $T\approx 1$ MeV 
is frozen into the expanding soup: $n/p \approx 0.20-0.25$.
You all know that neutrons outside of an atomic nucleus
are unstable particles and decay with a half-life of about 15 minutes.
But before that happens there is a possible escape route:
$$n+p \leftrightarrow D+\gamma;$$
that is to say, a neutron can combine with a proton to make a deuterium
nucleus and a photon.  However, so long as the mean energy of particles
and photons is greater than the binding energy of deuterium, about
86 Kev, the inverse reaction happens as well;  as soon as a deuterium
nucleus is formed it is photo-dissociated.  This means that
it is impossible to build up a 
significant abundance of deuterium until the temperature of the Universe
has fallen below 86 KeV or, looking back at eq. 4.4, until the Universe
has become older than about 2.5 minutes.  Then all of the remaining
neutrons are rapidly processed into deuterium.
\begin{figure}
\begin{center}
\includegraphics[height=10cm]{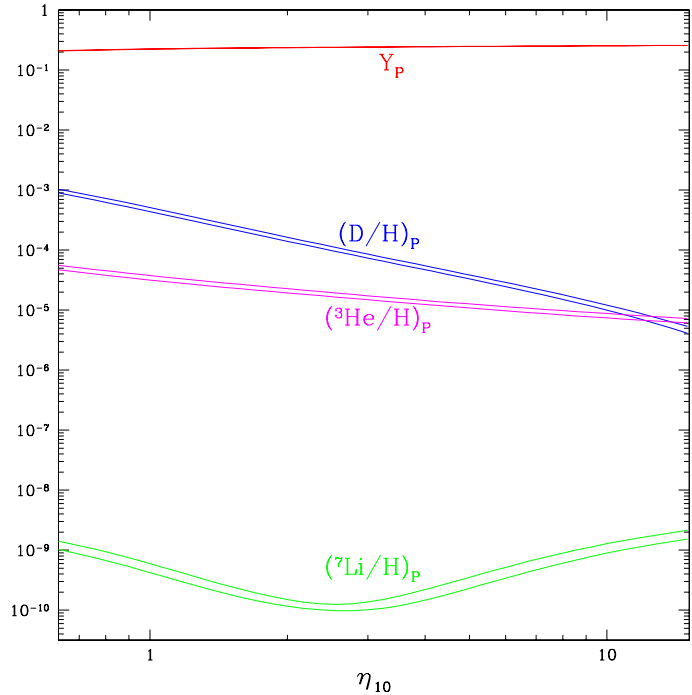}
\caption{The predicted abundances of the light isotopes as a function
of $\eta$ \cite{steig}.  
Here $Y_p$ is the predicted mass fraction of helium and is based upon the
assumption of three neutrino types.  The widths of the bands show the 
theoretical uncertainty.}
\end{center}
\end{figure}
\begin{figure}
\begin{center}
\includegraphics[height=10cm]{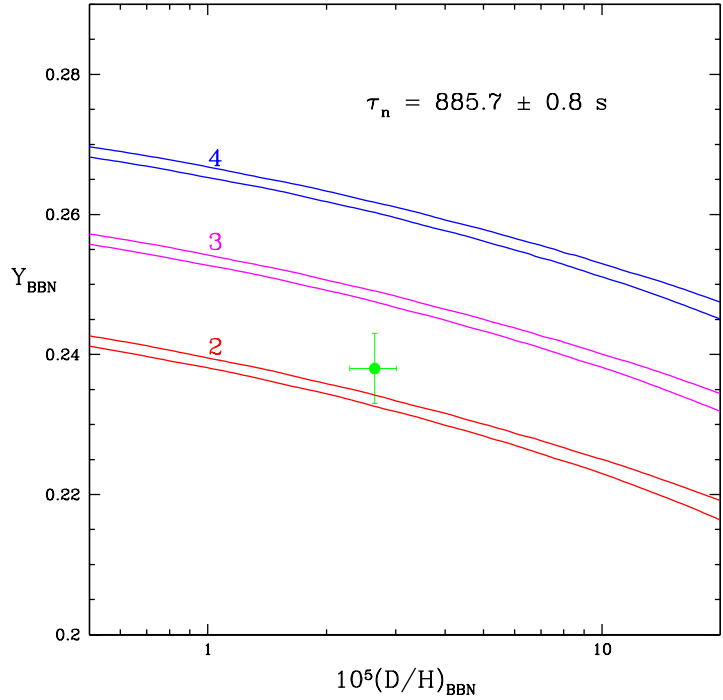}
\caption{The predicted abundance, $Y_p$, of helium (the mass fraction)
as a function of the predicted
deuterium abundance for two,three, and four neutrino types \cite{steig}.  
The point with error bars is the observed abundances of helium and deuterium.}
\end{center}
\end{figure}
But the deuterium doesn't stay around for long either.  Given the 
temperature and particle
densities prevailing at this epoch, there are a series
of two-body reactions by which two deuterons combine
to make He$^4$ and trace amounts of lithium and He$^3$.
These reactions occur at a rate which depends upon the overall abundance
of baryons, the ratio of baryons to photons: 
$$\eta = n_b/n_\gamma = 274\,\Omega_b h^2 \times 10^{-10}\eqno(4.5)$$
So essentially all neutrons which survive until T = 86 KeV become locked
up in He$^4$.  Therefore, the primordial abundance of helium depends
primarily upon the expansion rate of the Universe:  the faster the expansion
(due, say, to more neutrino types or to a larger constant of gravity) the
more helium.  The abundance of remaining deuterium, however, depends upon
the abundance of baryons, $\eta$:  the higher $\eta$ the less deuterium.
This is why it is sometimes said \cite{steig} that the 
abundance of primordial 
helium is a good chronometer (it measures the expansion rate), while the
abundance of deuterium is a good baryometer (it measures $\Omega_b$).
This is evident in Figs.\ 1 and 2 where we see first the predicted
abundances of various light isotopes as a function
of $\eta$, and secondly, the predicted abundance of He vs. that of
deuterium for two, three and four neutrino types.

The determination of primordial abundances is not a straightforward
matter because the abundance of these elements evolves due to processes
occurring within stars (``astration'').  In general, the abundance of
helium increases (hydrogen is processed to helium providing the primary
energy source for stars), while deuterium is destroyed by the same process.
This means that astronomers, when trying to estimate primordial abundances
of deuterium or helium, must try to find pristine, unprocessed material, in 
so far as possible.  One way to find unprocessed material is to look back
at early times, or large redshift, before the baryonic material has been
recycled through generations of stars.  This can be done with
quasar absorption line systems, where several groups of observers
have been attempting to 
identify very shallow absorption lines of deuterium at the same redshift
as the much stronger hydrogen Lyman alpha absorption line systems 
\cite{btyt,omea,petbow,kirk}.
It is a difficult observation requiring the largest telescopes;  the lines 
identified with deuterium might be mis-identified weak hydrogen or 
metal lines (incidentally, for an astronomer, any element heavier than
helium is a metal).  Taking the results of various groups at
face value, the weighted mean value 
\cite{steig} is
D/H $\approx 2.6\pm 0.3 \times 10^{-5}$.  Looking back at Fig. 1, we
see that this would correspond to $\eta = 6.1 \pm 0.6 \times 10^{-10}$
or $\Omega_b h^2 = 0.022\pm 0.003$.  

A word of caution is necessary here:
the values for the deuterium abundance determined by the different groups
scatter by more than a factor of two, which is considerably larger than 
the quoted statistical errors ($\approx 25\%$).  
This indicates that significant systematic effects are present.  But 
it is noteworthy that the angular power spectrum of the CMB anisotropies
also yields an estimate of the baryon abundance; this is encoded 
in the ratio of the amplitudes of the second to first peak.  The value is
$\Omega_b h^2 = 0.024 \pm 0.001$.  In other words, the two determinations
agree to within their errors.  This is quite remarkable considering that
the first determination involves nuclear processes occurring within the
first three minutes of the Big Bang, and the second involves 
oscillations of a photon-baryon plasma on an enormous scale
when the Universe is about 
500,000 years old.  If this is a coincidence, it is truly an astounding
one.

So much for the baryometer, but what about the chronometer-- helium?
Again astronomers are obliged to look for unprocessed material in order
to estimate the primordial abundance.  The technique of looking at 
quasar absorption line systems doesn't work for helium because the 
absorption lines from the ground state are far in the ultraviolet-- about
600 $\AA$ for neutral helium and, more likely, 300 $\AA$ from singly
ionized helium.  This is 
well beyond the Lyman limit of hydrogen, where the radiation from the
background quasar is effectively absorbed \cite{jak}.
Here the technique is to look for
He emission lines from HII regions (ionized gas around hot
stars) in nearby galaxies and compare to the hydrogen emission lines.
But how does one know that the gas is unprocessed?  The clue is in the fact
that stars not only process hydrogen into helium, but they also, in the
late stages of their evolution, synthesize heavier elements (metals) in
their interiors.  Therefore the abundance of heavier elements, like silicon,
is an indicator of how much nuclear processing the ionized gas has
undergone.  It is observed that the He abundance is correlated with the
metal abundance;  so the goal is to find HII regions with as low a
metal abundance as possible, and then extrapolate this empirical 
correlation to zero metal abundance \cite{olst,izthu}.  The answer turns out 
to be He/H $\approx$ 0.24, which is shown by the point with error bars in
Fig. 2.

This value is embarrassingly low, given the observed deuterium abundance.
It is obviously more consistent with an expansion rate provided by
only two neutrino types rather than three, but we know that there are
certainly three types.  Possible reasons for this apparent anomaly are:

1) Bad astronomy:  There are unresolved systematic errors
in determination of the relative He abundance in HII regions indicated
by the fact that the results of different groups differ by more than
the quoted statistical errors \cite{steig}.
The derivation of the helium to hydrogen ratio from the observed
He$^+$/H$^+$ ratio requires some understanding of 
the structure of the HII regions.  If there are relatively
cool ionizing stars ($T<35000$ K) spatially separated from the hotter stars,
there may be relatively less He$^+$ associated with a given
abundance of 
H$^+$.  Lines of other elements need to be observed to estimate the
excitation temperature;  it is a complex problem.

2) New neutrino physics:  There may be an asymmetry between neutrinos
and anti-neutrinos (something like the baryon- antibaryon asymmetry which
provides us with the observed Universe).  This would manifest itself
as a chemical potential in the Boltzmann equation giving different
equilibrium ratios of the various neutrino species \cite{barg}.

3) New gravitational physics:  any change in the gravitational interaction
which is effective at early epochs (braneworld effects?) could have
a pronounced effect on nucleosynthesis.  For example,  a lower effective 
constant of gravity would yield a lower expansion rate and
a lower He abundance.  The standard minimal braneworld correction term,
proportional to the square of the density \cite{brane}, 
goes in the wrong direction.

It is unclear if the low helium abundance is a serious problem for 
the standard Big Bang.
But it is clear that the agreement of the implied baryon
abundance with the CMB determination is an impressive success, and
strongly supports the assertion that the Hot Big Bang is the
correct model for the pre-recombination Universe.

\section{ The post-recombination Universe: determination of H$_o$ and
 t$_o$}

Certainly the most basic of the cosmological parameters is the present
expansion rate, \ho, because this sets the scale of the Universe.
Until a few years ago, there was a factor of two uncertainty in \ho;  with
two separate groups claiming two distinct values, one near 
50 km ${\rm s^{-1}{Mpc}^{-1}}$ and the other nearer 100 km 
${\rm s^{-1}{Mpc}^{-1}}$, and the errors quoted
by both groups were much smaller than this factor of two difference.  This
points out a problem which is common in observational cosmology (or indeed,
astronomy in general).  Often the indicated statistical errors give the
impression of great precision, whereas the true uncertainty is dominated
by poorly understood or unknown systematic effects.  That was true in
the Hubble constant controversy, and there is no less reason to think
that this problem is absent in modern results.  I will return to this point
several times below.

The great leap forward in determination of \ho  came with the 
Hubble Space Telescope
(HST) program on the distance scale.  Here a particular kind of 
variable stars-- Cepheid variables-- were observed in twenty nearby spiral
galaxies.  Cepheids exhibit periodic variations in luminosity by a factor
of two on timescales of 2-40 days.  There is a well-determined empirical
correlation between the period of Cepheids and their mean luminosity-- 
the longer the period the higher the luminosity.  Of course, this period-
luminosity relation must be calibrated by observing Cepheids in some object
with a distance known by other techniques and this remains a source of
systematic uncertainty.  But putting this problem aside, the Hubble
Space telescope measured the periods and the apparent magnitudes, without
confusion from adjacent bright stars, of a number of Cephieds in each of
these relatively
nearby galaxies, which yielded a distance determination (eq.\ 3.9).
These galaxies are generally too close (less than 15 Mpc) to sample 
the pure Hubble flow--  the Hubble flow on these scales is contaminated by
random motion of the galaxies and systematic cosmic flows-- 
but these determinations do permit a 
calibration of other secondary distance indicators which reach further out,
such as supernovae type Ia (SNIa) and the Tully-Fisher relation (the 
observed
tight correlation between the rotation velocities of a spiral galaxies and
their luminosities).  After an enormous amount of work by a number of very
competent astronomers \cite{freed}, the answer turned out to be
$h =0.72 \pm .10$

As I mentioned there is the known systematic uncertainty of calibrating
the period-luminosity relation, but there are other possible systematic 
effects that are less well-understood:  
How can we be certain that the period-luminosity relation for Cepheids
is the same in all galaxies?  For example, is this relation 
affected by
the concentration of elements heavier than helium (the 
metallicity)?  In view of such potential problems, other more direct
physical methods, which by-pass the traditional ``distance ladder''
are of interest.  Chief among these is the Sunyaev-Zeldovich (S-Z) effect
which is relevant to clusters of galaxies \cite{sz}.  The baryonic mass of 
clusters of galaxies is primarily in the form of hot gas, which typically
exceeds the mass in the visible galaxies by more than a factor of two.  
This gas
has a temperature between $10^7$ and $10^8$ K (i.e., the sound speed
is comparable to the one-dimensional velocity dispersion of the galaxies)
and is detected by satellite X-ray telescopes with detectors in the range
of several KeV.  The S-Z effect is a small change in the intensity of
the CMB in the direction of such clusters due to Compton scattering of
CMB photons by thermal electrons (classical electron scattering would,
of course, produce no intensity change).  Basically, CMB photons are
moved from the Rayleigh-Jeans part of the black body spectrum to the
Wien part, so the effect is observable as a spectral distortion of
the black body spectrum in the range of 100 to 300 GHz.  It is a
small effect (on the order of 0.4 milli Kelvin) but still 5 to 10
times larger than the intrinsic anisotropies in the CMB.

By measuring the amplitude of the S-Z effect one determines an
optical depth
$$\tau = \sigma n_el \eqno(5.1)$$
where $\sigma$ is the frequency dependent cross section, $l$ is the path
length, and $n_e$ is the electron density.
Because these same clusters emit X-rays via thermal bremsstrahlung, we may
also determine, from the observed X-ray intensity, an emission measure:
$$E = {n_e}^2 l \eqno(5.2)$$
Here we have two equations for two unknowns, $n_e$ and $l$.
(This is simplifying the actual calculation because $n_e$ is a 
function of radius in the cluster.)  Knowing $l$ and the angular 
diameter of the cluster $\theta$ we can then calculate the angular size
distance to the cluster via eq.\ 3.8.  Hence, the Hubble parameter is given
by $H_o = v/D_A$ where v is the observed recession velocity of the cluster.
All of this assumes that the clusters have a spherical shape on average,
so the method needs to be applied to a number of clusters.
Even so biases are possible if clusters have more typically
a prolate shape or an oblate shape, or if the X-ray emitting gas is
clumpy.  Overall, for a number of clusters
\cite{carls}
the answer turns out to be $h = 0.6$-- somewhat smaller than
the HST distance ladder method, but the systematic uncertainties remain
large.

A second direct method relies on time delays in gravitational lenses
\cite{refs}.
Occasionally, a distant quasar (the source) is lensed by an intervening
galaxy (the lens) into multiple images;  that is to say, we observe two
or more images of the same background object separated typically by
one or two seconds of arc.  This means that there are
two or more distinct null geodesics connecting us to the quasar with two
or more different light travel times.  Now a number of these quasars are 
intrinsically variable over time scales of days or months (not periodic 
but irregular variables).  Therefore, in two distinct images
we should observe the flux variations track each other with a time
delay.  This measured delay is proportional to the ratio
$D_lD_s/D_{ls}$ where these are the angular size distances to the lens,
the source, and the lens to the source.  Since this ratio is proportional
to ${H_o}^{-1}$, the measured time delay, when combined with a mass model for
the lens (the main source of uncertainty in the method), provides a
determination of the Hubble parameter.  
This method, applied to several lenses \cite{wilms,koopf}, 
again tends to yield
a value of $h$ that is somewhat smaller than the HST value, i.e.,
$\approx 0.6$.
In a recent summary \cite{koshe} it is claimed that, for four cases
where the lens is an isolated galaxy, the result is $h = 0.48\pm .03$,
if the overall mass distribution in each case can be represented
by a singular isothermal sphere.  On the other hand, in a well-observed
lens where the mass distribution is constrained by observations of
stellar velocity dispersion \cite{koopf2}, the implied value of $h$ is
$0.75^{+.07}_{-.06}$.  Such supplementary observations are important
because the essential uncertainty with this technique is in
the adopted  mass model of the lens.

It is probably safe to say that $h\approx 0.7$, with an uncertainty of
0.10 and perhaps a slight bias toward lower values, but the story is not 
over as S-Z and gravitational lens determinations continue to improve.
This is of considerable interest because the best fit to the CMB
anisotropies observed by WMAP implies that $h=0.72\pm .05$ 
in perfect agreement
with the HST result.  With the S-Z effect and lenses, there remains
the possibility of a contradiction.

With $h = .70$, we find a Hubble time of $t_H = 14$ Gyr.  Now in
FRW cosmology, the age of the Universe is $t_o = ft_H$ where $f$ is a
number depending upon the cosmological model.  For an Einstein-de Sitter
Universe (i.e., $\Omega_k = 0$, $\Omega_Q = 0$, $\Omega_m = 1$)
$f=2/3$ which means that $t_o = 9.1$ Gyr.  For an empty negatively curved
Universe, $f=1$ which means that the age is the Hubble time.  Generally,
models with a dominant vacuum energy density ($\Omega_Q \approx 1$,
$w\approx -1$)
are older ($f\ge 1$) and for the concordance model, $f=0.94$.
Therefore, independent determinations of the age of the Universe
are an important consistency test of the cosmology. 

It is reasonable to expect that the Universe should be older than the
oldest stars it contains, so if we can measure the ages of the oldest
stars, we have, at least, a lower limit on the age of the Universe.
Globular star clusters are old stellar systems in the halo of our
own galaxy;  these systems are distributed in a
roughly spherical region around the galactic disk and have low abundances
of heavy elements suggesting they were formed before most of the stars
in the disk.  If one can measure the luminosity, $L_u$, of the most luminous
un-evolved stars in a globular cluster (that is, stars still burning
hydrogen in their cores), then one may estimate the age.  That is
because this luminosity is correlated with age:  a higher $L_u$ means
a younger cluster.  Up to five years ago, this method yielded globular
cluster ages of $t_{gc} \approx 14\pm 2$ Gyr, which, combined with the
Hubble parameter discussed above, would be in direct contradiction with
the Einstein-de Sitter $\Omega_m=1$ Universe.  But about ten years ago
the Hipparchus satellite began to return accurate parallaxes for thousands
of relatively nearby stars which led to a recalibration of the 
entire distance scale.  Distances outside the solar system 
increased by about 10\% (in fact, the entire Universe suddenly
grew by this same factor leading to a decrease in the HST value for the
Hubble parameter).  This meant that the globular
clusters were further away, that $L_u$ was 20\% larger,
and the clusters were correspondingly younger:
$t_{gc}\approx 11.5\pm 1.3$ Gyr.  If we assume that the Universe is
about 1 Gyr older than the globular clusters, then the age of the
Universe becomes $12.5 \pm 2$ Gyr \cite{chab} which is almost consistent 
with the Einstein-de Sitter Universe.  At least there is no longer any 
compelling time scale argument for a non-zero vacuum energy 
density, $\Omega_Q>0$.  The value of accurate basic astronomical
data (and what is more basic than stellar positions?)  
should never be underestimated.

A second method for determining the ages of stars is familiar to all
physicists, and that is radioactive dating.  This has been done recently
by observations of a U$^{238}$ line in a metal-poor galactic star (an old
star).  Although the iron abundance in this star is only 1/800 that of 
the sun, the abundances of a group of rare earth metals known as 
r-process elements are
enhanced. The r-process is rapid neutron absorption onto iron nuclei
(rapid compared to the timescale for subsequent $\beta$ decay) which
contributes to certain abundance peaks in the periodic table and which
occurs in explosive events like supernovae.  This means that this old
star was formed from gas contaminated by an even older supernova event;
i.e. the uranium was deposited at a definite time in the past.
Now U$^{238}$ is unstable with a half life of 4.5 Gyr which makes it 
an ideal probe on cosmological times scales.  All we have to do is 
compare the observed abundance of $U^{238}$ to that of a stable
r-process element (in this case osmium), with what is expected from
the r-process.  The answer for the age of this star (or more accurately,
the SN which contaminated the gas out of which the star formed) is
12.5 $\pm 3$ Gyr, which is completely consistent with the globular cluster
ages \cite{cay}.

If we take $0.6 <h<0.7$, and $9.5\, {\rm Gyr}< t_o < 15.5\, {\rm Gyr}$ 
this implies
that $0.59<H_ot_o<1.1$.  This is consistent with a wide range of
FRW cosmologies from Einstein-de Sitter to the concordance model.
That is to say, independent measurements of \ho and $t_o$ are not yet
precise enough to stand as a confirmation or contradiction to
the WMAP result.

\section{Looking for discordance: the classical tests}

\subsection{ The angular size test}

The first of the classical cosmological tests we will consider is the
angular size test.  Here one measures the angular size of a standard
meter stick (hopefully) as a function of redshift;  different FRW cosmologies
make different predictions, but basically, for all FRW models $\theta(z)$
first decreases as $1/z$ (as would be expected in a Euclidean universe) 
and then increases with z.  This is because the angular size distance
is given by $D_A = r/(1+z)$ but the radial comoving coordinate approaches
a finite value as $z\rightarrow \infty$.  The angular size distance reaches
a maximum at a redshift between 1 and 2 and then decreases again.

When giant radio galaxies at large redshift were discovered in the 1960's
there was considerable optimism that these could be used as an angular size 
cosmological probe.  Radio galaxies typically have a double-lobe
structure with the radio emitting lobes straddling the visible
galaxy; these lobes
can extend hundreds of kpc beyond the visible object.  Such a linear
structure may be oriented at any angle to the observer's line-of-sight,
so one needs to measure the angular sizes of a number of radio galaxies
in a given redshift bin and only consider the largest ones, i.e.,  those
likely to be nearly perpendicular to the line-of-sight.

The result of all this work was disappointing.  It appeared that the angular
size of radio sources kept decreasing with redshift just as one would expect
for a pure Euclidean universe \cite{mile}.  The obvious problem,
that plagues all classical tests, is that of evolution.  Very likely, these
radio galaxies are not standard meter sticks at all, but that they were
actually smaller at earlier epochs than now.  This would be expected, because
such objects are thought to result from jets of relativistic particles
ejected from the nucleus of the parent galaxy in opposite directions.
The jets progress through the surrounding intergalactic medium at a 
rate determined by the density of that medium, which, of course, was 
higher at larger redshift.

But there is another class of radio sources that would be less 
susceptible to such environmental effects:  the compact
radio sources.  These are objects, on a scale of milli-arc-seconds, 
typically associated with distant quasars,
that are observed with radio interferometers having global baselines.
The morphology is that of a linear jet with lengths typically less than 
30 or 40 pc, so these would presumably be emission from the jets of
relativistic particles deep in the galactic nucleus near the central
engine producing them.  The intergalactic medium, and its cosmological
evolution, would be expected to have no effect here \cite{kell}.
\begin{figure}
\begin{center}
\includegraphics[height=8cm]{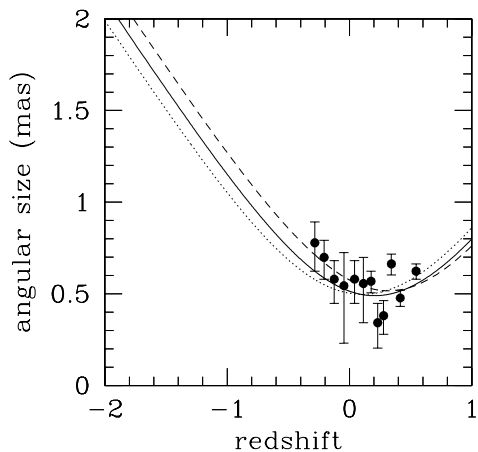}
\caption{The median angular size vs. redshift (log-log plots) for
145 compact radio
sources in 12 redshift bins.  The curves
are the three flat cosmological models: dashed, $\Omega_\Lambda = 0.9$;
solid, $\Omega_\Lambda = 0.7$ (concordance), dotted, $\Omega_\Lambda 
= 0.1$.  The physical size of the sources (20-40 pc) has been chosen
for the best fit}
\end{center}
\end{figure}

The result of plotting the median angular size of about 150
of these sources as a function of redshift is shown on a log-log plot
in Fig.\ 3  \cite{gurv}.  Also shown are the predicted relations 
for three flat
cosmologies ($\Omega_k=0$) with $\Omega_m = $ 0.9, 0.3, 0.1, the 
remainder being in a cosmological constant (the middle curve is
the concordance model).  In each case the linear size of the compact 
radio sources was chosen to achieve the best fit to the data.

It is evident that the general property of FRW models (that the angular
size of a standard meter stick should begin to increase again beyond
a redshift of about 1.5) is present in this data.  However, no statistical
test or maximum likelihood analysis is necessary to see that all three models
fit the data equally well.  This is basically an imprecise cosmological test
and cannot be improved, particularly considering that these objects may
also evolve in some unknown way with cosmic time.  Looking at the figure, one
may notice that measurement of angular sizes for just a few objects at
lower redshift might help distinguish between models.  However, there are
very few such objects at lower redshift, and these have a much lower
intrinsic radio power than those near redshift one.  It is dangerous to
include these objects on such a plot because they are probably of a very
different class.

\subsection{ The modern angular size test:  CMB-ology}

Although it is not my purpose here to discuss the CMB anisotropies,
it is necessary to say a few words on the preferred angular
scale of the longest wavelength acoustic oscillations, the  ``first peak'', 
because this is now the primary evidence
for a flat Universe ($\Omega_k=0$).  In Fig.\ 4  we see again the now
very familiar plot of the angular power spectrum of anisotropies as
observed by WMAP \cite{page} (in my opinion, of all the WMAP papers,
this reference provides the clearest discussion of the
physics behind the peak amplitudes and positions).
The solid line is the concordance model-- 
not a fit, but just the predicted angular power spectrum 
(via CMBFAST \cite{sz96}) from the $\Omega_m =
0.3$, $\Omega_\Lambda = 0.7$ model Universe with an optical depth
of $\tau \approx 0.17$ to the surface of last scattering.  
I must admit that the agreement is impressive. 
\begin{figure}
\begin{center}
\includegraphics[height=8cm]{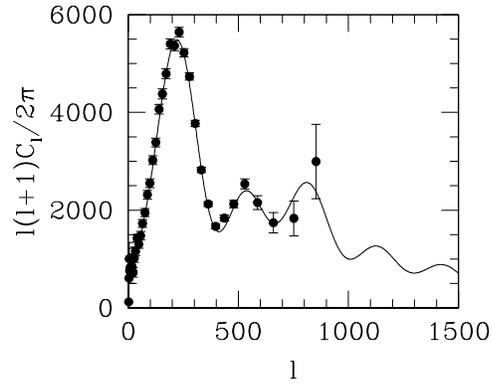}
\caption{The angular power spectrum of CMB anisotropies observed by
WMAP \cite{page}.  The solid line is not a fit but the is the
concordance model proposed earlier \cite{osst95}}
\end{center}
\end{figure}

I remind you that the harmonic index on the horizontal axis is related
to angular scale as
$$ l\approx \pi/\theta\eqno(6.1)$$
so the first peak, at $l\approx 220$, would correspond to an angular scale
of about one degree.  I also remind you that the first peak corresponds
to those density inhomogeneities which entered the horizon sometime before
decoupling (at $z=1000$);  enough before so that they have had time to
collapse to maximum compression (or expand to maximum rarefaction) just
at the moment of hydrogen recombination.  Therefore, the linear scale of these
inhomogeneities is very nearly given by the sound horizon at decoupling,
that is $$l_h \approx ct_{dec}/\sqrt{3} \eqno(6.2)$$
where $t_{dec}$ is the age of the Universe at decoupling.  
\begin{figure}
\begin{center}
\includegraphics[height=6cm]{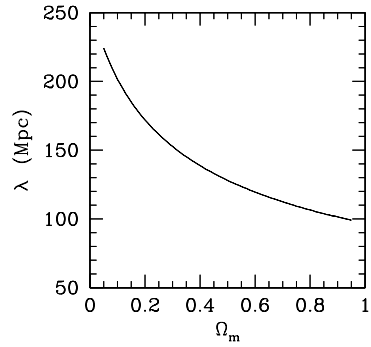}
\caption{The comoving linear scale of the perturbation corresponding 
to the first peak as a function of $\Omega_m$}
\end{center}
\end{figure}
\begin{figure}
\begin{center}
\includegraphics[height=6cm]{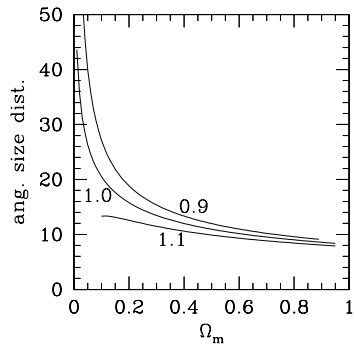}
\caption{The angular size distance (Gpc) to the last scattering surface
($z=1000$) as a function of $\Omega_m$ for various values of
$\Omega_{tot}$}
\end{center}
\end{figure}

So one might say, the test is simple: we have a known linear scale $l_h$
which corresponds to an observed angular scale ($\theta \approx 0.014$ rad)
so we can determine the geometry of the Universe.  It is not quite so
simple because the linear scale, $l_h$ depends, via $t_{dec}$ on the
matter content of the Universe ($\Omega_m$);  basically, the
larger $\Omega_m$, the sooner matter dominates the expansion, and the earlier
decoupling with a correspondingly smaller $l_h$.  This comoving linear scale
is shown in Fig.\ 5  as a function of $\Omega_m$ ($\Omega_\Lambda$ hardly 
matters here, because the vacuum energy density which dominates today has
no effect at the epoch of decoupling).  Another complication is that 
the angular size distance to the surface of last scattering not only 
depends upon the geometry, but also upon the expansion history.  This is
evident in Fig.\ 6 which shows the comoving angular size distance (in Gpc)
to the surface of last scattering as a function of $\Omega_m$ for three
values of $\Omega_{tot}=\Omega_m+\Omega_\Lambda$ (i.e., $\Omega_k = 
1-\Omega_{tot}$).  Note that the comoving angular size distance, $D_A(1+z)$,
is the same as the radial comoving coordinate $r$.

\begin{figure}
\begin{center}
\includegraphics[height=6cm]{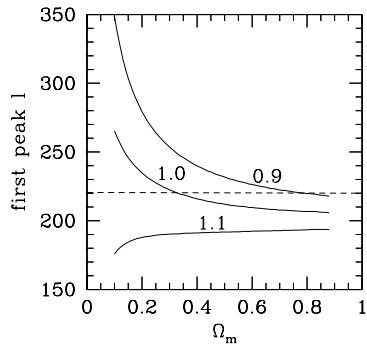}
\caption{The harmonic index expected for the first peak as a function
of $\Omega_m$ for various values of $\Omega_{tot}$.}
\end{center}
\end{figure}

We can combine Figs.\ 5 and 6  to plot 
the expected angular size (or harmonic index) of the first peak as
a function of $\Omega_m$ and $\Omega_{tot}$, and this is shown in
Fig.\ 7 with the dashed line giving the observed $l$ of the first peak.  
We see that
a model with $\Omega_{tot}\geq 1.1$  (a closed universe) is 
clearly ruled out, but it would be possible to have an open model with
$\Omega_{tot} = 0.9$ and $\Omega_m = 0.8$ from the position of the first
peak alone; the predicted peak amplitude, however, would be about
40\% too low.  The bottom line of all of this is that the {\it position} 
of the
first peak does not uniquely define the geometry of the Universe because
of a degeneracy with $\Omega_m$ (I haven't mentioned the degeneracy with
$h$ taken here to be 0.72).  
To determine whether or not we live in a flat Universe we need an
independent handle on $\Omega_m$ and that is provided, in WMAP data, by
the amplitudes of the first two peaks (the more non-baryonic matter, the
deeper the forming potential wells, and the lower the amplitudes).
From this it is found that $\Omega_m \approx 0.3$, and from Fig.\ 7 we see
that the model Universe should be near flat ($\Omega_{tot} \approx 1.0$).
Of course if the Universe is near flat with $\Omega_m = 0.3$ then the
rest must be in dark energy;  this is the indirect evidence from
the CMB anisotropies for dark energy.  

I just add here that the observed peak amplitudes (given the optical
depth to $z=1000$ determined from WMAP polarization results \cite{kogut}),
is taken now as definitive evidence for CDM.  However,
alternative physics which affects the amplitude
and positions of peaks (e.g. \cite{maart} could weaken this conclusion,
as well as affect the derived cosmological parameters.  Even taking
the peak amplitudes as {\it prima facie} evidence for the existence of
cold dark matter, it is
only evidence for CDM at the epoch of recombination ($z=1000$)
and not in the present Universe.  To address the cosmic coincidence
problem, models have been suggested in which dark matter transmutes
into dark energy (e.g. \cite{bass}).

Now I turn to the direct evidence for dark energy.

\subsection{ The flux-redshift test:  Supernovae Ia}

Type I supernovae are thought to be nuclear explosions of carbon/oxygen
white dwarfs in binary systems.  The white dwarf (a stellar remnant 
supported by the degenerate pressure of electrons) accretes matter from
an evolving companion and its mass increases toward the Chandrasekhar
limit of about 1.4 M$_\odot$ (this is the mass above which the degenerate
electrons become relativistic and the white dwarf unstable).  Near this
limit there is a nuclear detonation in the core in which 
carbon (or oxygen) is converted to iron.  A nuclear flame propagates to
the exterior and blows the white dwarf apart (there are alternative
models but this is the favored scenario \cite{branch}).

These events are seen in both young and old stellar populations; for
example, they are observed in the spiral arms of spiral galaxies where
there is active star formation at present, as well as in elliptical 
galaxies where vigorous star formation apparently ceased many Gyr ago.
Locally, there appears to be no difference in the properties of SNIa arising
in these two different populations, which is important because at 
large redshift the stellar population is certainly younger.

The peak luminosity of SNIa is about $10^{10}$ L$_\odot$ which is comparable
to that of a galaxy.  The characteristic decay time is about one month which,
in the more distant objects, is seen to be stretched by 1+z as expected.
The light curve has a characteristic form and the spectra contain
no hydrogen lines, so given reasonable photometric and spectroscopic
observations, they are easy to identify as SNIa as opposed to type II
supernovae; these are thought to be explosions of young massive stars
and have a much larger dispersion in peak luminosity \cite{leib}.

The value of SNIa as cosmological probes arises from the high peak luminosity
as well as the observational evidence (locally) that this peak
luminosity is the sought-after standard candle.  In fact, the absolute
magnitude, at peak, varies by about 0.5 magnitudes which corresponds
to a 50\%-60\% variation in luminosity; this, on the face of it, 
would make them
fairly useless as standard candles.  However, the peak luminosity appears
to be well-correlated with decay time:  the larger L$_{peak}$, the slower
the decay.  There are various ways of quantifying this effect
\cite{leib}, such as
$$M_B \approx 0.8(\Delta m_{15} -1.1) -19.5 \eqno(6.3)$$
where $M_B$ is the peak absolute magnitude and
$\Delta\,\,m_{15}$ is the observed change in apparent magnitude 15 days
after the peak \cite{ham}.  This is an empirical relationship, and there is
no consensus about the theoretical explanation, but, when this correction is
applied it appears that $\Delta L_{peak} < 20\%$.  If true, this means
that SNIa are candles that are standard enough to distinguish
between cosmological models at $z\approx 0.5$.

In a given galaxy, supernovae are rare events (on a human time scale, that is),
with one or two such explosions per century.  But if thousands of galaxies
can be surveyed on a regular and frequent basis, then it is possible to
observe several events per year over a range of redshift.  About 10 years
ago two groups began such ambitious programs \cite{pereal99,garn};  
the results have
been fantastically fruitful and have led to a major paradigm shift.

\begin{figure}
\begin{center}
\includegraphics[height=8cm]{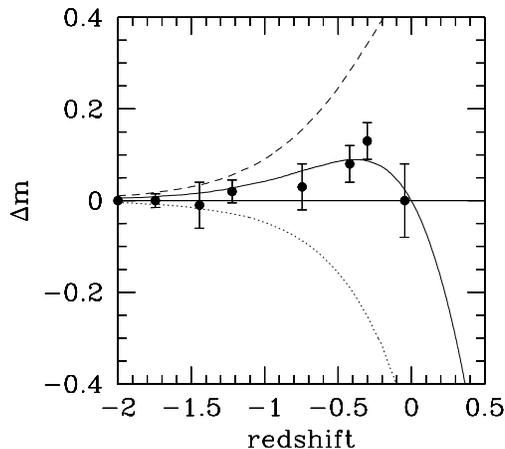}
\caption{The Hubble diagram for SNIa normalized to an empty non-accelerating
Universe.  The points are binned median values for 230 supernovae \cite{tonry}
The curves show the predictions for three flat ($\Omega_{tot}
=0$) cosmological models:  The dashed line is the model dominated by a 
cosmological constant ($\Omega_{\Lambda}=0.9$), the solid curve is the
concordance model ($\Omega_{\Lambda}=0.7$), and the dotted curve is the
matter dominated model ($\Omega_{\Lambda}=0.1$).}
\end{center}
\end{figure}

The most recent results are summarized in \cite{tonry}:  at present,
about 230 SNIa have been observed out to $z=1.2$.  The bottom line is that
SNIa are 10\% to 20\% fainter at $z\approx 0.5$ than would be expected in
an empty ($\Omega_{tot}=0$) non-accelerating Universe.  But, significantly, 
at $z\ge 1$ the supernovae appear to become brighter again relative to the
non-accelerating case;  this should happen in the concordance model at 
about this redshift because it is here that the cosmological constant
term in the Friedmann equation (eq.\ 3.7  ) first begins to dominate over the
matter term.  This result is shown in Fig.\ 8 which is a plot of
the median $\Delta m$, the observed deviation from the non-accelerating
case, in various redshift bins as a function of redshift (i.e., the
horizontal line at $\Delta m = 0$ corresponds to the empty universe).
The solid curves show the prediction for various flat ($\Omega_{tot}=1$)
models with the value of the cosmological term indicated.  It is evident
that models dominated by a cosmological term or by matter are inconsistent
with the observations at extremely high levels of significance, while the
concordance model agrees quite well with the observations.

It is also evident from the figure that the significance of the effect is not
large, perhaps 3 or 4 $\sigma$ (quite a low level of significance on which
to base a paradigm shift).  When all the observed supernovae are included
on this plot, it is quite a messy looking scatter with a minimum 
$\chi^2$ per degree of freedom (for flat models) which is greater than one.
Moreover the positive result depends entirely upon the empirical
peak luminosity-decay rate relationship and, of course, upon the assumption
that this relation does not evolve.  So, before we become too enthusiastic
we must think about possible systematic effects and how these might affect
the conclusions.  These effects include:

1) Dust:  It might be that supernovae in distant galaxies are more
(or less) dimmed by dust than local supernovae.  But normal dust,
with particle sizes comparable to the wavelength of light, not only
dims but also reddens (for the same reason, Rayleigh scattering, that
sunsets are red).  This is quantified by the so-called color excess.
Remember I said that astronomers measure the color of an object by
its B-V color index (the logarithm of a flux ratio).  The color excess
is defined as $$E(B-V) = (B-V)_{obs}-(B-V)_{int}\eqno(6.4)$$
where $obs$ means the observed color index and $int$ means the intrinsic
color index (the color the object would have with no reddening).  In our
own galaxy it is empirically the case that the magnitudes of absorption
is proportional to this color excess, i.e., $$A_V=R_V E(B-V)\eqno(6.5)$$
where $R_V$ is roughly constant and depends upon average grain properties.
So assuming that the dust in distant galaxies is similar to the dust
in our own, it should be possible to estimate and correct for the
dust obscuration.  Significantly \cite{pereal99}, it appears that there
is no difference between E(B-V) for local and distant supernovae.
This implies that the distant events are not more or less obscured than
the local ones.

2) Grey dust:  It is conceivable (but unlikely) that intergalactic space 
contains dust particles which are significantly larger than the wavelength
of light.  Such particles would dim but not redden the distant supernovae 
and so would be undetectable by the method described above \cite{aghai}.
It is here that
the very high redshift supernovae ($z>1$) play an important role.  If this
is the cause of the apparent dimming we might expect that the supernovae
would not become brighter again at higher redshift.

3) Evolution:  It is possible that the properties of these events
may have evolved
with cosmic time.  As I mentioned above, the SN exploding at high
redshift come from a systematically younger stellar population than the 
objects observed locally.  Moreover, the abundance of metals was smaller
in the earlier Universe than now;  this evolving composition, by changing
the opacity in the outer layers or the composition of the fuel itself
could lead to a systematic evolution in peak luminosity.  Here it is 
important to look for observational differences between local and distant
supernovae, and there seem to be no significant differences in most respects,
the spectrum or the light curve.  There is, however, a suggestion that 
distant supernovae are intrinsically bluer than nearby objects \cite{leib}.  
If this effect is verified, then it could not only 
point to a systematic difference
in the objects themselves, but could also have lead to an underestimate of 
the
degree of reddening in the distant SN.  It is difficult, in general, to
eliminate the possibility that the events themselves were different
in the past and that this could mimic the effect of a cosmological constant
\cite{drell};  a deeper theoretical understanding of the SNIa process
is required in order to realistically access this possibility.

4) Sample evolution:  The sample of SN selected at large redshift may 
differ from the nearby sample that is used, for example, to calibrate
the peak luminosity-decline rate correlation.  There does appear to
be an absence, at large redshift, of SN with very slowly declining
light curves-- which is to say, very luminous SN that are seen 
locally.  Perhaps a class of more luminous objects is missing in
the more distant Universe due to the fact that these SN emerge from
a systematically younger stellar population.  One would hope that
the luminosity-decline rate correlation would correct for this effect,
assuming, of course, that this relation itself does not evolve.

5) Selection biases:  There is a dispersion in the luminosity-decline
rate relationship, and in a flux-limited sample, one tends to select
the higher luminosity objects.  Astronomers call this sort of bias 
the ``Malmquist effect'' and it is always present in such
observational data.  Naively,  one would expect such a bias to
lead to an
underestimate of the true luminosity, and, therefore an underestimate
luminosity distance;  the bias actually diminishes the apparent 
acceleration.     
But there is another effect which is more difficult to access:  The
most distant supernovae are being observed in the UV of their own
rest frame.  SNIa are highly non-uniform in the UV, and K-corrections
are uncertain.  This could introduce systematic errors at
the level of a few hundredths of a magnitudes \cite{tonry}.

We see that there are a number of systematic effects that could bias
these results.  A maximum likelihood analysis over the entire
sample \cite{tonry}, confirms earlier results that the 
confidence contours in $\Omega_m$-$\Omega_\Lambda$ space are stretched
along a line $\Omega_\Lambda = 1.4\Omega_m + 0.35$ and that the actual
best fit is provided by a model with $\Omega_m\approx 0.7$ and
$\Omega_\Lambda \approx 1.3$-- not the concordance model.  Of course,
if we add the condition that $\Omega_{tot} = 1$ (a flat Universe) then
the preferred model becomes the concordance model.  In \cite{tonry} it
is  suggested that this apparent deviation is due to the appearance of 
one or more of
the systematic effects discussed above near $z=1$ at the level of 0.04 
magnitudes.
 
The result that SNIa are systematically dimmer near $z=0.5$ than expected
in a non-accelerating Universe is robust.
At the very least it can be claimed with reasonable certainty that the 
Universe is not decelerating at present.  However, given the probable presence
of systematic uncertainties at the level of a few hundredths of a magnitude,
it is difficult to constrain the equation of state ($w$) of the dark
energy or its evolution ($dw/dt$) until these effects are better understood.
I will just mention that lines of constant age, $t_oH_o$, are 
almost parallel to the best fit line in the $\Omega_m$-$\Omega_\Lambda$
plane mentioned above.  This then gives a fairly tight constraint on
the age in Hubble times \cite{tonry}; i.e. $t_oH_o = 0.96\pm 0.4$, 
which is consistent with the WMAP result.  In 
a near flat Universe this rules out the dominance of matter and requires
a dark energy term.

\subsection {Number counts of faint galaxies}

The final classical test I will discuss is that of number counts of 
distant objects-- what radio astronomers call the log(N)-log(S) test.
Basically one counts the number of galaxies N brighter than
a certain flux limit S.  If we lived in a static Euclidean universe,
then the number of galaxies out to distance R would be $N\propto R^3$
but the flux is related to R as $S\propto R^{-2}$.  This implies
that $N\propto S^{-3/2}$ or $log(N) = -3/2\,log(S) + const = 0.6m + const.$
where m is the magnitude corresponding to the flux S. 
\begin{figure}
\begin{center}
\includegraphics[height=8cm]{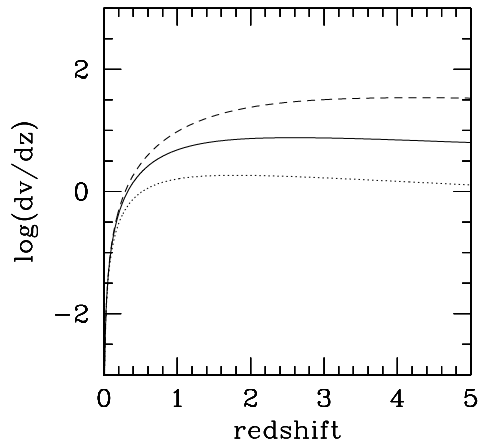}
\caption{The log of the incremental volume per incremental redshift
(in units of the Hubble volume)
as a function of redshift for the three flat cosmological models}
\end{center}
\end{figure}

But we do not live in a static Euclidean universe; we live in
an evolving universe with a non-Euclidean geometry where the differential
number counts probe dV(z), the comoving volume as a function of redshift.
In Fig.\ 9 we see log(dV/dz) as a function of redshift
for three different($\Omega_{tot} = 1$) cosmological models:  
the matter dominated
Universe, the cosmological constant dominated Universe, and the concordance
model.  For small z, dV/dz increases as $z^2$ for all models
as would be expected in a Euclidean Universe, but by redshift one, the
models are obviously diverging, with the models dominated by a cosmological
constant having a larger comoving incremental volume.  Therefore if we can
observe faint galaxies extending out to a redshift of one or two, we might
expect number counts to provide a cosmological probe. 

There is a long history of counting objects as a function of flux or 
redshift.  Although cosmological conclusions have been drawn
(see, e.g.\ \cite{fuku}), the overall consensus is that this is not a 
very good test
because the galaxy population evolves strongly with redshift.
Galaxies evolve because stars evolve.  In the past, the stellar populations
were younger and contained relatively more massive, luminous stars.
Therefore we expect galaxies to be more luminous at higher redshift. 
It is also possible that the density of galaxies evolves because
of merging, as would be consistent with the preferred model of
hierarchical structure formation in the Universe.

The distribution of galaxies by redshift can be used, to some extent, to
break this degeneracy between evolution and cosmology.  If we can measure
the redshifts of galaxies with infrared magnitudes between 23 and 26, for
example, that distribution will be skewed toward higher redshift if there
is more luminosity evolution.

I have recently reconsidered the number counts of the faint galaxies in
the Hubble Deep Fields, north and south \cite{willeal,caseal}.  
These are two separate small
patches of empty sky observed with the Hubble Space Telescope down to
a very low flux limit-- about $m_I$ = 30 (the I band is a far red
filter centered around 8000 angstroms). 
The differential number counts are shown by the solid
round points in Fig.\ 10 where ground based number counts at fainter 
magnitudes are also shown by the starred points.  

\begin{figure}
\begin{center}
\includegraphics[height=8cm]{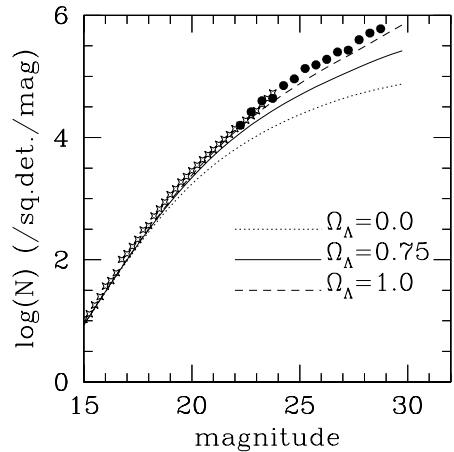}
\caption{The solid points are the faint galaxy number counts from the
Hubble Deep Fields (north and south \cite {willeal,caseal}) 
and the star shaped points are the
number counts from ground based data.  The curves are the no-evolution
predictions from three flat cosmological models.}
\end{center}
\end{figure}

For this same sample of galaxies, there are also
estimates of the redshifts based upon the galaxy colors-- so called 
photometric redshifts \cite{fern}.  In order to calculate the expected number
counts and redshift distribution one must have some idea of the form of
the luminosity function-- the distribution of galaxies by redshift.
Here, like everyone else, I have have assumed that this form is
given by the Schechter function \cite{schect}:
$$N(L)dL = N_o(L/L_*)^{-\alpha} exp(-L/L_*)dL\eqno(6.6)$$
which is characterized by three parameters: $\alpha$, a power law at low
luminosities, $L_*$ a break-point above which the number of galaxies 
rapidly decreases, and $N_o$ a normalization.  I take this form 
because the overall galaxy distribution by luminosity at low redshifts is 
well fit by such a law \cite{blant}, so I am assuming that at least the form
of the luminosity function does not evolve with redshift.  

\begin{figure}
\begin{center}
\includegraphics[height=8cm]{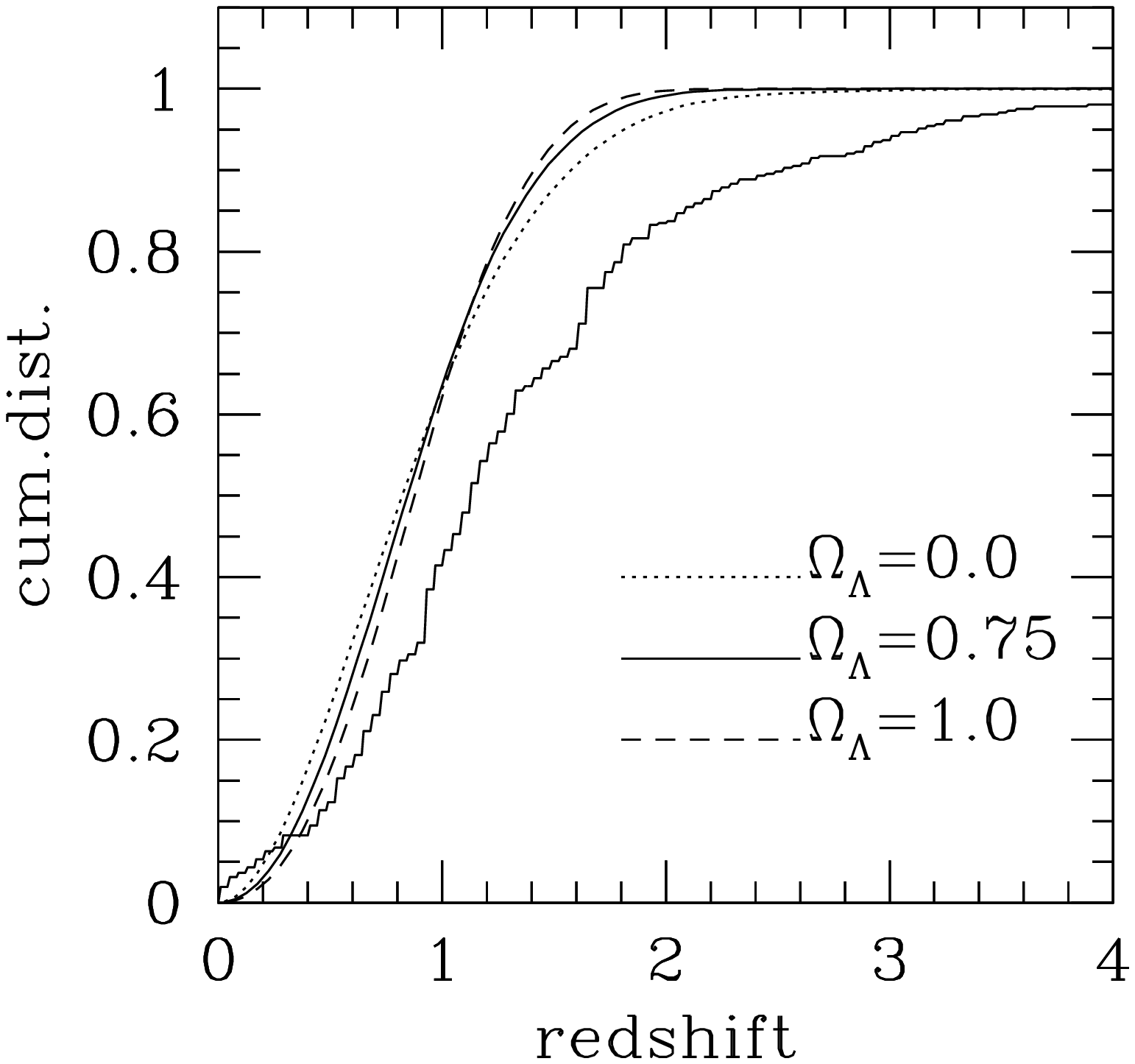}
\caption{The cumulative redshift distribution for galaxies between
apparent I-band magnitudes of 23 and 26 (photometric redshifts from 
\cite{fern}).
The curves are the predicted no-evolution distributions for the three
cosmological models.}
\end{center}
\end{figure}

But when I consider
faint galaxies at high redshift in a particular band I have to be careful to 
apply the K-correction mentioned above; that is, I must correct the
observed flux in that band to the rest frame.  Making this correction
\cite{pogg}, but
assuming no luminosity or density evolution, I find the differential
number counts
appropriate to our three flat cosmological models shown by the indicated
curves in Fig.\ 10.  We see that the predicted number counts all fall short
of the observed counts, but that the cosmological constant dominated model
comes closest to matching the observations.  However, 
the distribution by redshift of HDF galaxies between I-band magnitudes of
22 and 26 is shown in Fig.\ 11 (this is obviously the cumulative 
distribution).  Here we
see that all three models seriously fail to match the observed distribution,
in the sense that the predicted mean redshift is much too small.
\begin{figure}
\begin{center}
\includegraphics[height=8cm]{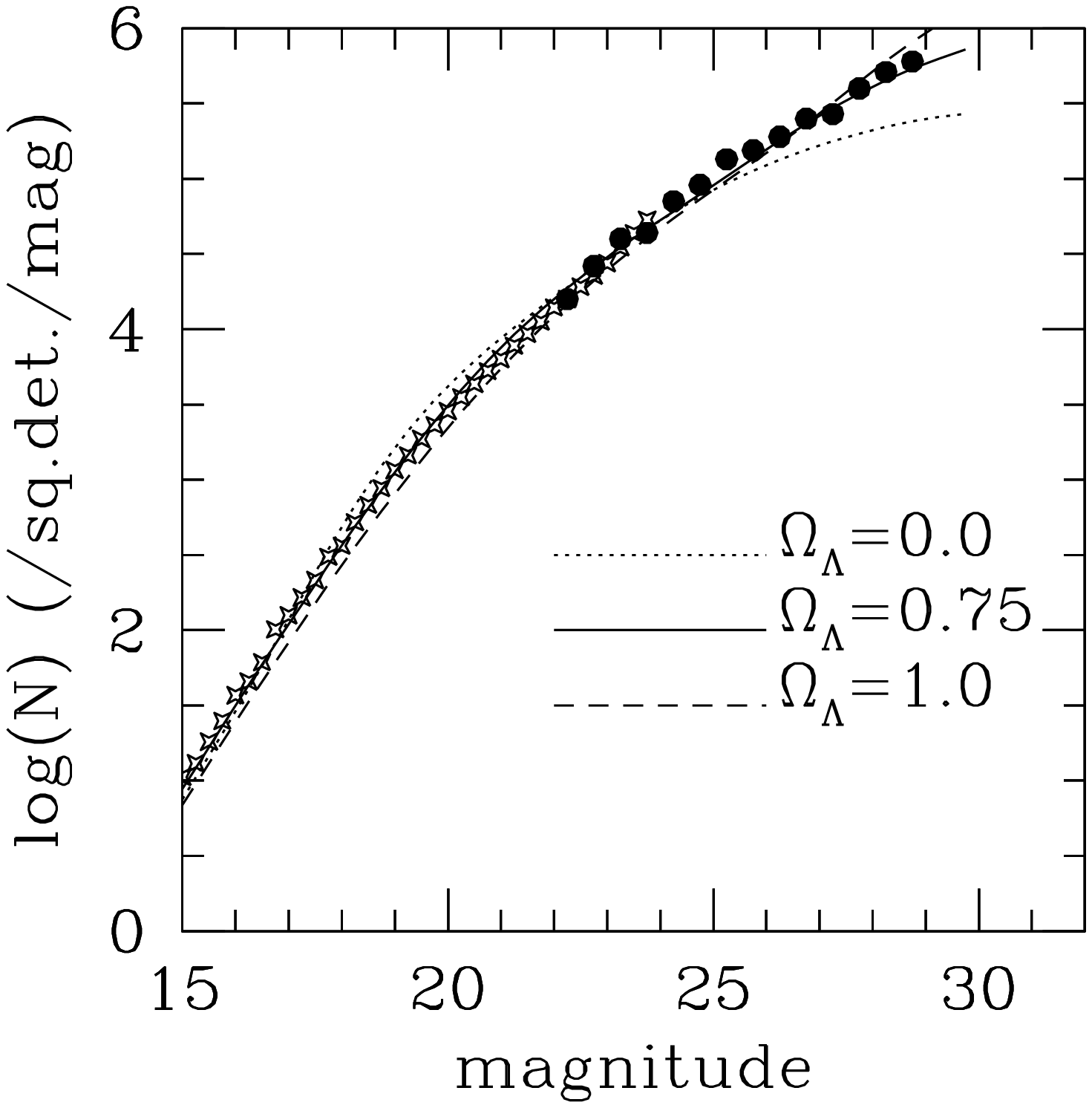}
\caption{As in Fig.\ 10 above the observed galaxy number counts and the
predictions for the cosmological models with luminosity evolution
sufficient to explain the number counts.}
\end{center}
\end{figure}

This problem could obviously be solved by evolution.  If galaxies are
brighter in the past, as expected, then we would expect to shift this
distribution toward higher redshifts.  One can conceive of very
complicated evolution schemes, involving initial bursts of star
formation with or without continuing star formation,  but it would
seem desirable to keep
the model as simple as possible;  let's take a ``minimalist'' model for 
galaxy evolution.  A simple one parameter scheme with 
the luminosity brightening proportional to the look-back time squared, i.e.,
every galaxy brightens as 
$$\Delta M_I = q\,{(H_ot_{lb})}^2 \eqno(6.7)$$
where $q$ is the free parameter, can give a reasonable
match to evolution models for galaxies \cite{pogg}.  
(we also assume that all galaxies are the same-- they are not divided into
separate morphological classes).  I choose the value of $q$ such that
the predicted redshift distribution most closely matches the observed
distribution for all three models, and the results are shown in Fig.\ 12.

\begin{figure}
\begin{center}
\includegraphics[height=8cm]{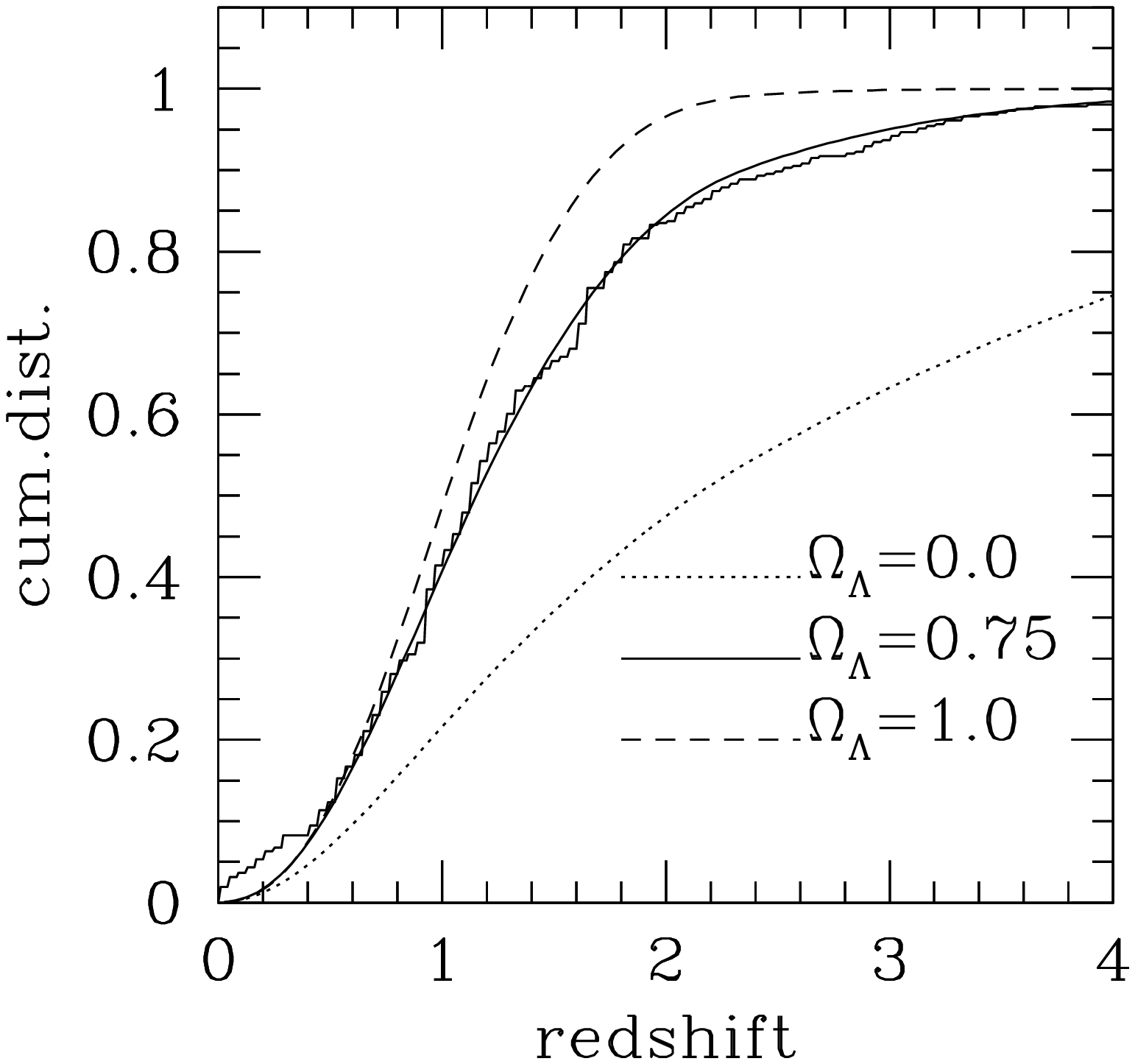}
\caption{The cumulative redshift distribution for galaxies between
apparent i-band magnitudes of 22 and 26 (photometric redshifts from 
\cite{fern}).
The curves are the predicted distributions for the three
cosmological models with evolution sufficient to explain the number
counts.}
\end{center}
\end{figure}

The required values of $q$  (in magnitudes per ${t_H}^2$)
for the three cosmological models are: $q=2.0$ ($\Omega_\Lambda = 1.0$),
$q=3.0$ ($\Omega_\Lambda= 0.7$), and $q=11.0$ ($\Omega_\Lambda=0.0$).
Obviously, the matter-dominated model requires the most evolution, and
with this simple evolution scheme, cannot be made to perfectly match
the observed distribution by redshift (this in itself is not definitive
because one could always devise more complicated schemes which would work).
For the concordance model, the required evolution would be about
two magnitudes out to $z = 3$.

For these same evolutionary models, that is, with 
evolution sufficient to match the number counts, the
predicted redshift distributions are shown in Fig.\ 13.  
Here we see that the model dominated
by a cosmological constant predicts too many low redshift galaxies, 
the matter
dominated model predicts too few, and the model that works perfectly is
very close to the concordance model!  Preforming this operation for
a number of flat models with variable $\Omega_\Lambda$, I find that
$0.59 < \Omega_\Lambda < 0.71$ to 90\% confidence.

Now there are too many assumptions and simplifications to make this 
definitive.  The only point I want to make is that faint galaxy number
counts and redshift distributions are completely consistent with the
concordance model when one considers the simplest minimalist model for
pure luminosity evolution.  One may certainly conclude that number counts 
provide no contradiction
to the generally accepted cosmological model of the Universe (to my 
disappointment).

\section{ Conclusions}

In these lectures I have been looking for discord, but have not found it.
The classical tests return results for cosmological parameters that are
consistent with but considerably less precise than those implied by the 
CMB anisotropies,  given the usual assumptions.
It is fair to say that the numbers characterizing the concordance
model,
$\Omega_m \approx 0.3$, $\Omega_\Lambda \approx 0.7$ are robust {\it in the
context of the framework of FRW cosmology}.  It is, in fact, the peculiar
composition of the Universe embodied by these numbers which calls that
framework into question.

Rather small changes in the assumptions underlying pure FRW cosmology 
(with only an evolving vacuum energy density 
in addition to more familiar fluids) 
can make a difference.  For example, allowing $w=-0.6$ brings
the number counts and z-distribution of faint galaxies into agreement with
a Universe strongly dominated by dark energy 
($\Omega_Q = 0.9$).  The same also true of the
high-z supernovae observations \cite{tonry}).  Allowing
a small component of correlated iso-curvature initial perturbations,
as expected in braneworld cosmologies, can affect the amplitudes and
positions of
the peaks in the angular power spectrum of the CMB anisotropies
\cite{maart}, and therefore the derived cosmological
parameters.

But even more drastic changes have been suggested.  Certain
braneworld scenarios, for example, in which 4-D gravity is induced
on the brane \cite{dvali}
imply that gravity is modified at large
scale where gravitons begin to leak into the bulk \cite{deff}.
It is possible that the observed acceleration is due to such modifications
and not to dark energy.  More ad hoc modifications of General Relativity
\cite{ceal} have also been proposed because of a general unease
with dark energy-- proposals whereby gravity is modified in the limit
of small curvature scalar.  My own opinion is that we should also feel
uneasy with the mysterious non-baryonic cold dark matter, because the
only evidence for its existence, at present, is its gravitational influence;
when the theory of gravity is modified to eliminate dark energy, it might
also be found that the need for dark matter vanishes.  

In general,  more attention
is being given to so-called infrared modifications of gravity
(e.g. \cite{arkan}), and this is a positive
development.  High energy modifications, that affect the evolution of
the early Universe, are, as we have seen, strongly constrained by
considerations of primordial nucleosynthesis (now, in combination
with the CMB results).  It is more likely that modifications play
a role in the late, post-recombination evolution of the Universe,
where the peculiarities of the concordance model suggest that they
are needed.  The fact
that the same rather un-natural values for the comparable densities of 
dark energy and matter 
keep emerging in different observational contexts may be calling
attention to erroneous underlying assumptions rather than to the
actual existence of these ``ethers''.  

Convergence toward a parameterized
cosmology is not, without deeper understanding, sufficient reason for 
triumphalism.  Rather, it should be a motivation to look 
more carefully at the possible systematic effects in the observations
and to question more critically the underlying assumptions of the models.
\parskip 5truemm
\parindent 0pt

I thank Rien van de Weygaert, Ole M\"oller, Moti Milgrom, Art Wolfe, 
Jacob Bekenstein, and Scott Trager 
for useful comments on the manuscript.  I also thank Gary Steigman,
Wendy Freedman, and Luis Ho for permission to use Figs.\ 1 and 2.
I am very grateful to the organizers of the Second Aegean Summer School
on the Early Universe, and especially, Lefteris Papantonopoulos, 
for all their work and for inviting me to the very pleasant island of Syros.

\end{document}